\documentclass[a4paper]{ESASPCS13Style}
\usepackage{epsfig}

\begin{document}

\title{Spot the differences: The X-ray spectrum 
of SU~Aur compared to TW~Hya}

\author{K.\ Smith\inst{1} \and M.\
  Audard\inst{2} \and M.\ G\"udel\inst{3} \and S.\ Skinner\inst{4}
\and R.\ Pallavicini\inst{5}} 
\institute{Max-Planck-Institut f\"ur Radioastronomie, Auf dem H\"ugel 69,
  D-53121 Bonn, Germany, \and
  Columbia Astrophysics Laboratory, Columbia University, Mail Code 5247, 
550 West 120th Street, New York, NY 10027, USA 
\and Paul Scherrer Institute, Villigen \& W\"urenlingen, 5232 Villigen PSI, 
Switzerland
\and Center for Astrophysics and Space Astronomy, University of Colorado, Boulder, CO 80309-0389, USA
\and Osservatorio Astronomico di Palermo, Piazza del Parlamento 1, 
90134 Palermo, Italy }

\maketitle 

\begin{abstract}

We present high-resolution Chandra HETGS X-ray spectra of the
classical T Tauri star SU Aur. The quiescent X-ray emission is
dominated by a 20--40~MK plasma, which contrasts strongly with the cool
3~MK plasma dominating the X-ray emission of the CTTS TW~Hya.  A large flare
occurred during the first half of our 100~ks observation, and we have
modelled the emitting plasma both during this flare and during the
apparently quiescent periods. During the flare, an extremely high
temperature plasma component (at least 60~MK) accounts for the bulk of
the emission. There is an indication of the presence of Fe~XXVI
emission at 1.78~\AA, which is maximally formed at 130~MK.

\keywords{Stars: X-rays -- Stars: activity -- Stars: individual: SU Aur --
  Stars: pre-main-sequence }
\end{abstract}

\section{Introduction}

The classical T Tauri stars (CTTS) are pre-main-sequence low-mass
objects which continue to accrete from circumstellar disks. Stellar
magnetic fields disrupt the inner disk and funnel the accreting matter
inward.  Because of the star-disk coupling, the CTTS generally rotate
more slowly than their diskless relatives, the weak-lined T Tauri
stars. This slower rotation may lead to lower levels of coronal activity.

The special nature of CTTS leads to a number of possible observable properties
in their X-ray emission, which might distinguish them from WTTS or from active
main sequence stars.  Differential rotation between star and inner disk could
lead to the winding up of field lines, reconnection, and hence large,
quasi-periodic flares. Such events may have been observed in the young object
YLW~15 (Tsuboi et al. 2000). The star-disk interaction zone is invoked in many
different models of jet launching and collimation. The accretion shock on the
stellar surface is expected to have a temperature of around $10^{6}$K
(0.1~keV), so cool plasma may dominate the emission measure
distribution. Material loaded onto the field lines from the disk could lead to
higher density plasma than seen in purely coronal sources, where the particles
are evaporated from the stellar chromosphere. Abundance anomalies, compared to
main-sequence dwarfs, might also be expected.

Many of these properties were seen in the spectrum of the nearby (57~pc) CTTS
TW~Hya. Kastner et al. (2002) observed this star with the {\em Chandra } HETGS
and found that the emission measure distribution was dominated by cool 3~MK
plasma, consistent with temperatures expected in an accretion shock. The
density sensitive He-like Ne~IX triplet at 13.45, 13.55 and 13.7~\AA\ implies
log $n_e$=12.75~cm$^{-3}$, a high density consistent with a typical CTTS
accretion rate of around $10^{-8} M_{\odot}$/yr (Gullbring et al. 2000). The
high density and emission measure imply a small emitting volume relative to
most coronal sources. The Ne abundance was found to be a factor of 2-3 higher
than solar, while Fe and O were deficient.  These results were also found by
Stelzer \& Schmitt (2004), who observed with {\em XMM-Newton} RGS. They
suggested that the abundance anomalies could be due to depletion of grain
forming metals in a disk. All these results are consistent with a picture of
X-ray emission from accretion shocks caused by dense material falling in from
a circumstellar disk.

It is important that spectra of more CTTS are obtained, to ascertain whether
this accretion picture is applicable to CTTS as a class.  Unfortunately, few
CTTS are bright enough to undertake high-resolution spectroscopy with {\em
Chandra} or XMM{\em-Newton} -- TW~Hya is itself accessible mainly due to its
proximity. SU~Aur is one of the X-ray-brightest CTTS and one of the few
feasible targets. We present here our {\em Chandra } HETG spectra of SU Aur
and compare them to the TW~Hya results and to results for other young stars.

\subsection{SU Aur}
\label{sec:suaur}

SU~Aur is a relatively massive CTTS, at about 2~$M_{\odot}$ with
spectral type G2~III. It has an estimated age of 4--5~Myr and lies in
the Taurus-Auriga complex at a distance of 152~pc. The accretion rate is
estimated to be around $10^{-8}$M$_{\odot}$yr$^{-1}$. {\em ASCA}
observations by Skinner \& Walter (1998) showed that the
differential emission measure distribution was dominated by hot
($\sim$20~MK) plasma. 

\section{Analysis}
\label{sec:anal}

\begin{figure}
\begin{center}
\epsfig{file=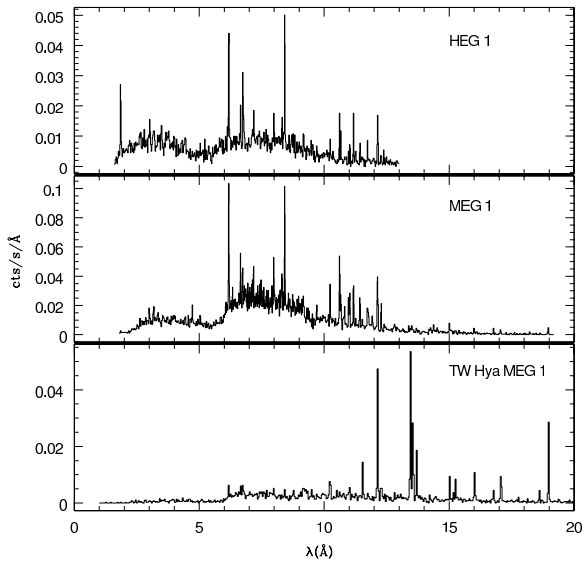,angle=0,width=8cm}
\caption{HEG (top) and MEG (middle) spectra for the complete
observation of SU Aur. The spectrum of TW~Hya is also shown (bottom panel).
\label{fig1}}
\end{center}
\end{figure}

Figure~\ref{fig1} shows the {\em Chandra} HEG and MEG first-order spectra.
Also shown is the MEG 1st order spectrum of TW~Hya.  The contrast between the
spectra of the two stars is immediately clear.  The SU~Aur spectrum shows the
presence of hot continuum emission below 5~\AA. Various moderate-temperature
lines such as Si~XIV (6.2~\AA, peak emissivity at 16~MK) and Mg~XII (8.4~\AA,
peak emissivity at 10~MK) are also strong in SU Aur, whilst they are weak or
absent for TW Hya. Low-temperature lines such as Ne~X at 12.1~\AA\ or O~VIII at
16~\AA\ are much stronger in the TW~Hya spectrum than for SU~Aur.

A further indicator of hot plasma in SU~Aur is the presence of Fe~XXV emission
at 1.85~\AA, with a peak emissivity at 60~MK, and Fe~XXVI at 1.78~\AA, with a
peak emissivity at 125~MK. These lines are evidence of a very hot plasma
component during the observation.

\begin{figure}[t]
\begin{center}
\epsfig{file=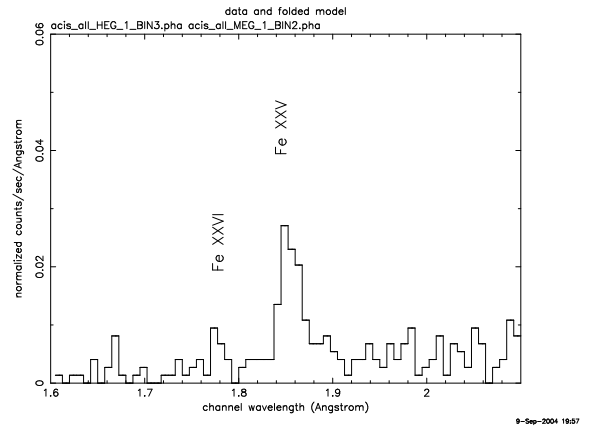,angle=270,width=8cm }
\caption{The short-wavelength part of the complete HEG spectrum, showing 
the Fe~XXV line at 1.85~\AA\ and the Fe XXVI line at 1.78~\AA. \label{fig2}}
\end{center}
\end{figure}

\begin{figure}[h]
\begin{center}
\epsfig{file=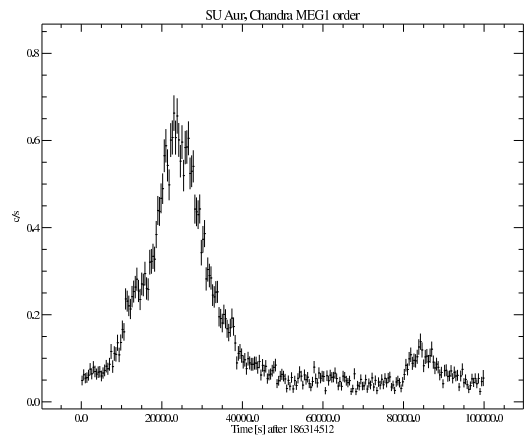,angle=90,width=8cm}
\caption{MEG lightcurve. An apparent 'shoulder' on the large flare
at around 17~ks may indicate that this flare is in fact a composite of at
least two successive flare events. \label{fig3}}
\end{center}
\end{figure}

\begin{figure}[h]
\begin{center}
\epsfig{file=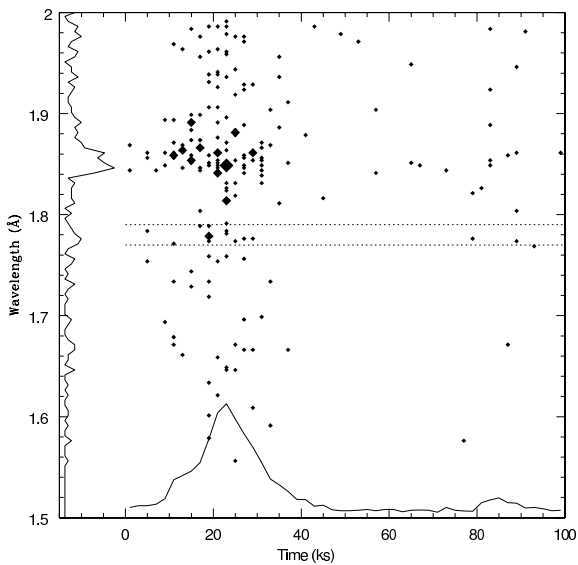,angle=0,width=8cm}
\caption{Scatterplot of arrival times (x-axis) of photons against 
wavelength (y-axis), for the portion of the spectrum 
below 2~\AA. The spectrum is broken down into 5~ks segments. 
Larger symbols represent the arrival of 2, 3 or 4 photons in the same 5~ks
segment. The lightcurve is plotted 
on the x-axis, and the spectrum on the y-axis. 
The dashed lines show the position of the Fe~XXVI line. \label{fig4}}
\end{center}
\end{figure}

\subsection{flaring behaviour}
\label{sec:flares}

The MEG 1st order lightcurve (Fig.~3) shows flaring behaviour in the early
part of the observation.  The large rise here may be caused by at least
two separate flares, the second one beginning as the first peaks around
T=13~ks -- producing a 'shoulder' in the flare profile at this time.  
The large flare or flares together radiate approximately $10^{36}$~erg/s.  A
further, smaller flare occurred later at around T=85~ks. These three components
are indicated on the plot, the first one also with an 
exponential decay. It is therefore sensible to split the analysis
between the quiescent and flaring emission, where signal-to-noise constraints
allow.  

One question which can be posed here is when the Fe~XXV, Fe~XXVI and
short wavelength continuum emission are detected. We obviously expect the bulk
of these photons to arrive during the flaring periods, when the plasma
component is hottest. In Fig.~4 we show a scatterplot of the arrival times of
the various photons in the short-wavelength part of the spectrum. 
From this figure it can be seen that {\em no} photons at wavelengths shorter
than 1.8~\AA\ arrive during the quiescent periods. The Fe~XXV line at 1.85~\AA\
is largely formed during the large flare with some contributions during the
later flare and during the quiescent periods. The suspected Fe~XXVI line is
formed only during the flaring periods, with the overwhelming majority of
photons arriving during the large early flare.

Models with three temperature components and photoelectric absorption were
used to fit the spectra. The background was negligable, and therefore was not
subtracted, allowing robust C statistics to be used for the fit. We
considered three separate time intervals when fitting models; the period from
5~ks to 22.5~ks, which is dominated by rising flare emission, from 22.5~ks to
50~ks, which is dominated by the flare decay phase, and a quiescent state
comprising three intervals, 0--5~ks, 50-80~ks and 90--100~ks.  In the case of
the quiescent emission, a two-temperature model was found to be sufficient to
fit the data.
\begin{figure}
\begin{center}
\epsfig{file=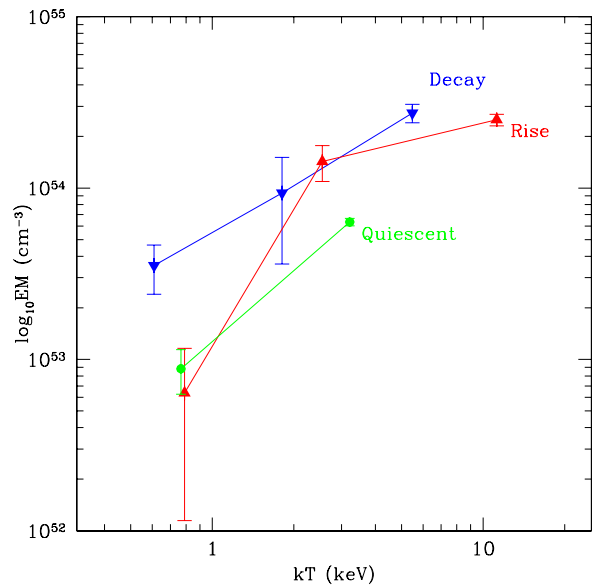,angle=0,width=8cm}
\caption{Emission measure and temperature of components of 3-temperature fits
to the rising (red, upward-pointing triangles) and decaying (blue,
downward-pointing triangles) emission and a 2-temperature fit to the quiescent
emission (green, squares). \label{fig5}}
\end{center}
\end{figure}
The rise phase is dominated by plasma at around 100~MK and 25~MK. The decay
phase is dominated by slightly cooler plasma at around 50~MK and 20~MK. The
quiescent plasma is dominated by plasma at around 30-40~MK.  These fits
further confirmed the strong presence of a very hot ($\ge$60~MK) plasma
component during the flaring periods, and a hot component ($\sim$40~MK) which
dominates {\em even during quiescence}. The emission measure and temperature
of each model component for each fit is shown in Figure~\ref{fig5}

\begin{figure}[h]
\begin{center}
\epsfig{file=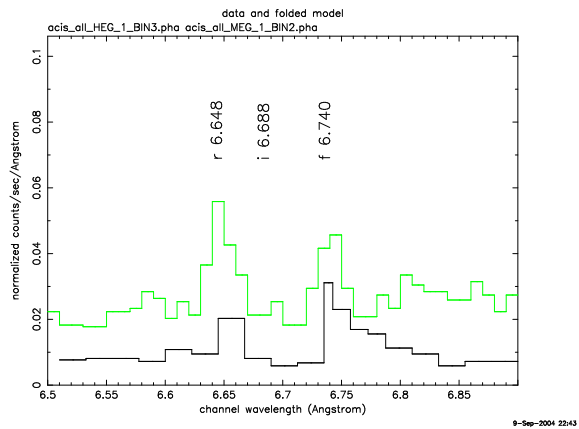,angle=270,width=8cm}
\caption{The Si~XIII He-like triplet at 6.6--6.8~\AA. The MEG spectrum is
shown in green (light grey in the black and white version) and the HEG
spectrum is in black. The resonance, intercombination and forbidden lines are
labelled. \label{fig6}}
\end{center}
\end{figure}

A full analysis of the density sensitive He-like triplets has not yet
been attempted. Triplets available in the HETGS spectra
include S~XV at 5-5.1~\AA, sensitive to densities between
$10^{13}-10^{15}$cm$^{-3}$, Si~XIII at 6.6-6.8~\AA\ 
($10^{13}-10^{15}$cm$^{-3}$),  Mg~XI at 9.2-9.3~\AA\ 
($10^{12}-10^{14}$cm$^{-3}$), and Ne~IX at  13.5-13.7~\AA\ 
($10^{11}-10^{13}$cm$^{-3}$). The accretion rate of 
$10^{-8}$M$_{\odot}$yr$^{-1}$ suggests a density of around 
$5 \times 10^{12}$cm$^{-3}$. Unfortunately, the Ne~IX and 
Mg~XI triplets, which are sensitive to such low densities, are 
extremely weak in our spectra. The strongest He-like triplet is 
Si~XIII, shown in Fig.~\ref{fig6}. It is clear that the forbidden line 
here is stronger than the intercombination line, suggesting that the 
density is less than $10^{13}$cm$^{-3}$.  

A preliminary investigation of the abundances of various metals relative to Fe
was made by fitting the entire time integrated spectrum.  The fitted abundance
of Fe was 0.89 solar during quiescence and 1.07 solar during the
flare. Abundance ratios relative to Fe were found to be Mg/Fe=0.85,
Si/Fe=0.89, S/Fe=0.41, O/Fe=0.98 and Ne/Fe=1.08. There is some evidence that
the abundance of oxygen rises during the flaring state (O/Fe=0.6 during
quiescence), whilst that of Ne falls (Ne/Fe=1.98 during quiescence).  A full
analysis is necessary before definitive conclusions can be drawn regarding the
abundances.

\section{Conclusions}
\label{sec:conc}

The high-resolution {\em Chandra} HETGS spectra we present here show clearly
the presence of hot (30~MK) plasma in the quiescent emission.  This is
augmented by the presence of an even hotter component (60+~MK) during the
large flares seen in the first half of the observation.  The presence of such
hot plasma is also indicated by the observation of Fe~XXVI at 1.78~\AA.  Very
hot plasma (up to 100 MK) in stellar flares is not unusual in active stars,
and has been detected directly in some cases through hard X-ray emission seen
with {\em Beppo}SAX 
(e.g. Franciosini et al. 2001 on UX Ari, Maggio et al. 2000 on AB Dor
and Favata \& Schmitt 1999 on Algol). This EM distribution does however make a
strong contrast to TW~Hya, the only CTTS previously studied at this spectral
resolution in the X-ray. Hot (45~MK) plasma is also reported elsewhere in
these proceedings for the CTTS RY Tau (Audard et al. 2004).  The appearance of
the Si~XIII forbidden line indicates a density less than $10^{13}$~cm$^{-3}$.
Although this density constraint is not very stringent, there is no compelling
evidence in our preliminary analysis of the SU~Aur spectrum for densities as
high as those inferred for TW~Hya by Kastner et al. (2002).
There is evidence from preliminary fitting that the Ne abundance may
be high compared to other metals during quiescence, whilst O may have a low
abundance (but the Fe abundance appears approximately solar). The Ne and O
abundances change towards solar during the flare.

In summary, SU~Aur provides at least one counterexample 
to the properties of TW~Hya. It may be that SU Aur is the unusual 
object -- the sensitivity limits of current high-resolution X-ray 
spectrometers introduce a strong selection bias -- but it would 
nevertheless be premature to adopt TW~Hya as the prototypical
classical T Tauri X-ray source at this stage. 

\begin{acknowledgements}

The authors would like to acknowledge support from SAO grants GO3-4015A
and G03-4015B, and also from the Swiss NSF (grant 20-66875.01).  

\end{acknowledgements}

\end{document}